\documentclass{amsart}
\usepackage{amsfonts}

\theoremstyle{plain}
\newtheorem{theorem}{Theorem}[section]

\theoremstyle{remark}
\newtheorem{remark}[theorem]{Remark}
\numberwithin{equation}{section}

\input{tcilatex}

\begin{document}

\thispagestyle{empty}

\begin{center}
{\footnotesize Available at: \texttt{http://publications.ictp.it}}\hfill
IC/2008/...\\[0pt]
\vspace{1cm} United Nations Educational, Scientific and Cultural Organization%
\\[0pt]
and\\[0pt]
International Atomic Energy Agency\\[0pt]
\medskip THE ABDUS SALAM INTERNATIONAL CENTRE FOR THEORETICAL PHYSICS\\[0pt]

\bigskip

\vspace{1.8cm}THE VACUUM STRUCTURE, SPECIAL RELATIVITY THEORY

AND QUANTUM MECHANICS REVISITED: \ A FIELD THEORY-NO-GEOMETRY APPROACH

\bigskip

\bigskip

\bigskip {\small \textit{The authors dedicate this article to one of the
mathematical and physical giants of  the  XX-th century - \\[0pt]
academician Prof. Nikolai N. Bogolubov in memory of his 100th Birthday with
great appreciation to his brilliant talent and impressive impact to  modern
nonlinear mathematics and quantum physics}}\\[0pt]
\bigskip

\bigskip \vspace{1.5cm} Nikolai N. Bogolubov (Jr.)\footnote{%
nikolai\_bogolubov@hotmail.com}\\[0pt]
\bigskip \textit{The V.A. Steklov Mathematical Institute of RAN, Moscow,
Russian Federation \\[0pt]
and \\[0pt]
The Abdus Salam International Centre for Theoretical Physics, Trieste, Italy,%
}\\[1em]
\smallskip Anatoliy K. Prykarpatsky\footnote{%
pryk.anat@ua.fm, prykanat@cybergal.com}\\[0pt]

\bigskip

\textit{The AGH University of Science and Technology, Krak\'{o}w 30-059,
Poland,\\[0pt]
and\\[0pt]
The Ivan Franko State Pedagogical University, Drogobych, Lviv region,
Ukraine \\[0pt]
} \smallskip and\\[1em]
Ufuk Taneri\footnote{%
ufuk.taneri@gmail.com}\\[0pt]

\bigskip

\textit{The Department of Applied Mathematics and Computer Science, Eastern
Mediterranean University EMU, Famagusta, North Cyprus \\[0pt]
and\\[0pt]
Kyrenia American University GAU, Institute of Graduate Studies, Kyrenia,
North Cyprus.}\\[0pt]
\end{center}

\baselineskip=18pt

\vfill

\begin{center}
MIRAMARE -- TRIESTE\\[0pt]
September 2008\\[0pt]
\end{center}

\vfill\newpage \setcounter{page}{1} \centerline{\bf Abstract}The main
fundamental principles characterizing the vacuum field structure are
formulated and the modeling of the related vacuum medium and charged point
particle dynamics by means of devised field theoretic tools are analyzed.
The Maxwell electrodynamic theory is revisited and newly derived from the
suggested vacuum field structure principles and the classical special
relativity theory relationship between the energy and the corresponding
point particle mass is revisited and newly obtained. The Lorentz force
expression with respect to arbitrary non-inertial reference frames is
revisited and discussed in detail, and some new interpretations of relations
between the special relativity theory and quantum mechanics are presented.
The famous quantum-mechanical Schr\"{o}dinger type equations for a
relativistic point particle in the external potential and magnetic fields
within the quasiclassical approximation as the Planck constant $\hbar
\rightarrow 0$ and the light velocity $c\rightarrow \infty $ are obtained.

\bigskip\ \newpage

\section{Introduction}

It is a generally accepted statement that no one physical theory can present
the absolutely true picture of \ Nature, assuming that there always exist
boundaries of its application, which are approved by experiments and new
experience data. This statement concerns, evidently, the relativity theory
which, as it was pondered by A. Einstein, should logically \ arise from both
the light velocity constance principle \ with respect to the inertial
reference frames and  the generalized equivalence principle with respect to
gravity and inertial masses. This theory, namely the special relativity
theory, proved to be very thoroughly confirmed by many \ nuclear physics \
experiments, which simultaneously became the nuclear energetics backgrounds,
widely used today worldwide.

Nevertheless, the nature of space-time and surrounding matter objects was
and persists to be \  one of the most intriguing and challenging problems
facing  mankind and, in particular,  natural scientists.  As we know,  one
of the most brilliant inventions in physics of 19-th century was the
combining of electricity and magnetism within the Faraday-Maxwell
electromagnetism theory. This theory explained the main physical laws of
light propagation in space-time and posed new questions concerning the
nature of vacuum. Nonetheless, almost \ all the attempts aiming to unveil
the real state of art of the vacuum problem appeared to be unsuccessful in
spite of new ideas suggested by Mach, Lorentz, Poincar\'{e}, Einstein and
some others physicists. \ Moreover, the non-usual way of treating the
space-time devised by Einstein, in reality, favored  eclipsing both its
nature and the related physical vacuum origin problems \cite%
{Fe,Ge,Ma,TW,Ba,PB,Ca}, reducing them to some physically unmotivated formal
mathematical principles and recipes, combined in the well-known special
relativity theory (SRT). \ The SRT appeared to be adapted \ only  to the
inertial reference systems and met with hard problems of the electromagnetic
Lorentz forces explanation and relationships between inertial and gravity
forces. The latter was artificially "dissolved" by means of the well-known
"equivalence principle"  owing to which the "inertial" mass of a material
object was postulated to coincide with its "gravity" mass. In contrast, E.
Mach suggested that any motion of a material point in the space-time, both
straight-linear and curvi-linear, can  also be only  relative. In
particular, as the inertion law is also relative, the measure of material
point inertion, that is its mass, should  also be relative and depend on
mutual interactions between all material bodies in the Universe. These E.
Mach's ideas  influenced \ so strongly A. Einstein that he dreamed of them
his whole life, trying to reconcile Mach's relativity principle with his own
approach to the general relativity theory. Nonetheless, despite the titanic
efforts of A. Einstein, he failed to include the Mach relativity principle
into his general relativity theory, since they appeared to be mutually
excluding each other.

Simultaneously, the vacuum origin as a problem almost completely disappeared
from the Einstein theory being replaced by the geometrization of \
space-time nature and all related physical phenomena. Meanwhile, the
impressive success of 20-th century quantum physics, especially  quantum
electrodynamics, have demonstrated clearly enough \cite%
{Fe1,So,BS,BjD,AB,Sch,B-B,Ba} that the vacuum polarization and
electron-positron annihilation phenomena make it possible to pose new
questions about  space-time and vacuum structures, and further to revisit
\cite{Br,Ca,Ma,Fa,BD,Lo,Fo1,Re,PB} the existing points of view on them.

As  well known, the classical mechanics uses the notion of "potential"
energy, being a scalar function of spatial variables, \ very important \ for
formulating dynamical equations, in spite of the fact that it is determined
up to arbitrary constant. \ One\ can also observe a \ similar situation  in
the classical electrodynamic theory, which effectively uses the notions of
scalar and potentials related to each other via the well-known Lorentz
compatibility gauge constraint \cite{BjD,AB,BS,Pa} and defined up to
suitable gauge transformations. These, in some sense, "different" potential
functions were later deeply reanalyzed within the classical Einstein
relativity theory by many physicists \cite%
{So,BjD,Fa,Br,Lo,Pa,Re,Fo1,BD,Ma,We,LK,L-L,Me1,Se} that gave rise to the
understanding of  their fundamental role in combining two great theories -
the electromagnetism and the gravity.

Moreover, new important problems arose owing to the famous Einstein
relationship between the internal energy and the velocity dependent mass,
belonging to a material particle. But, as it was mentioned by L. Brillouin
\cite{Br}, \ the relationship \ between the particle mass and its internal
energy takes into account  only the  kinetic energy, describing no mass
pertaining to the potential energy, which  makes the classical relativity
theory \ a not completely closed and, physically, not compatible theoretical
construction. And as  written by P.W. Bridgeman in \cite{Bri}, \ "... at
construction of his general relativity theory A. Einstein did not make use
of those lessons, which he had us taught himself, and of his deep
penetration, which he demonstrated us in his special relativity theory".

Below we try to unveil some nontrivial aspects of the real space-time and
vacuum origin problems, \ deeply related with the relativity theory and
electrodynamics, to derive, from the\ natural field theory principles, all
of the well-known Maxwell electromagnetism and special relativity theories
results, to show their relative or only visible coincidence with real
physical phenomena and to feature new perspectives facing the modern
fundamental physics.

Moreover, having further developed the field approach to the microscopic
vacuum structure, previously  suggested in \cite{Re} and accounted for in
\cite{PB}, we obtained, within the quasi-classical approximation, a new
derivation of main quantum mechanical relationships describing evolution of
microscopic particle systems, coinciding as $\hbar \rightarrow 0$ with those
devised at the beginning of the 20-th century by the great physicists Schr%
\"{o}dinger, Heisenberg and Dirac.

\section{The Maxwell electromagnetism theory: new look and interpretation}

We start from the following field theoretical model \cite{PB} of the
microscopic vacuum structure, considered as some physical reality imbedded
into the standard three-dimensional Euclidean space reference system marked
with three spatial coordinates $r\in \mathbb{R}^{3},$ endowed with the
standard scalar product $<\cdot ,\cdot >,$ and parameterized by means of the
scalar temporal parameter $t\in \mathbb{R}.$ \ We will describe the physical
vacuum matter endowing it with an everywhere enough smooth four-vector
potential function $(W,A):\mathbb{R}^{3}\times \mathbb{R}\rightarrow \mathbb{%
R}\times \mathbb{R}^{3},$ naturally related to light propagation properties.
The material objects, imbedded into the vacuum, we will model (classically
here) by means of the scalar charge density function $\ \rho :\mathbb{R}%
^{3}\times \mathbb{R}\rightarrow \mathbb{R}$ and the vector current density $%
\ J:\mathbb{R}^{3}\times \mathbb{R}\rightarrow \mathbb{R}^{3},$ being also
everywhere enough smooth functions.

\begin{enumerate}
\item The \textit{first} field theory principle regarding the vacuum we
accept is formulated as follows: the four-vector function $(W,A):\mathbb{R}%
^{3}\times \mathbb{R}\rightarrow \mathbb{R}\times \mathbb{R}^{3}$ satisfies
the standard Lorentz type continuity relationship%
\begin{equation}
\frac{1}{c}\frac{\partial W}{\partial t}+<\nabla ,A>=0,  \label{M1.1}
\end{equation}%
where, by definition, $\nabla :=\partial /\partial r$ is the usual gradient
operator.

\item The \textit{second} field theory principle we accept is a dynamical
relationship on the scalar potential component $W:\mathbb{R}^{3}\times
\mathbb{R}\rightarrow \mathbb{R}:$%
\begin{equation}
\frac{1}{c^{2}}\frac{\partial ^{2}W}{\partial t^{2}}-\nabla ^{2}W=\rho ,
\label{M1.2}
\end{equation}%
assuming  the linear law \ of the small vacuum uniform and isotropic
perturbation propagations in the space-time, understood here, evidently, as
a first (linear) approximation in the case of weak enough fields.

\item The \textit{third }principle is similar to the first one and means
simply the \  continuity condition for the density and current density
functions:%
\begin{equation}
\frac{\partial \rho }{\partial t}+<\nabla ,J>=0.  \label{M1.3}
\end{equation}
\end{enumerate}

We need to note here that the vacuum field perturbations velocity parameter $%
c>0,$ used above, coincides with the vacuum light velocity,  as we are
trying to derive successfully from these first principles the well-known
Maxwell electromagnetism field equations, to analyze the related Lorentz
forces and \ special relativity relationships. To do this, we first combine
equations (\ref{M1.1}) and (\ref{M1.2}):%
\begin{equation*}
\frac{1}{c^{2}}\frac{\partial ^{2}W}{\partial t^{2}}=-<\nabla ,\frac{1}{c}%
\frac{\partial A}{\partial t}>=<\nabla ,\nabla W>+\rho ,
\end{equation*}%
whence
\begin{equation}
<\nabla ,-\frac{1}{c}\frac{\partial A}{\partial t}-\nabla W>=\rho .
\label{M1.4}
\end{equation}%
Having put, by definition,%
\begin{equation}
E:=-\frac{1}{c}\frac{\partial A}{\partial t}-\nabla W,  \label{M1.5}
\end{equation}%
we obtain the first material Maxwell equation%
\begin{equation}
<\nabla ,E>=\rho   \label{M1.6}
\end{equation}%
for the electric field $E:\mathbb{R}^{3}\times \mathbb{R}\rightarrow \mathbb{%
R}^{3}.$ Having now applied the rotor-operation $\ \ \nabla \times $ \ to
expression (\ref{M1.5}) we obtain the first Maxwell field equation%
\begin{equation}
\frac{1}{c}\frac{\partial B}{\partial t}-\nabla \times E=0  \label{M1.7}
\end{equation}%
on the magnetic field vector function $B:\mathbb{R}^{3}\times \mathbb{R}%
\rightarrow \mathbb{R}^{3},$ defined as \ \
\begin{equation}
B:=\nabla \times A.  \label{M1.8}
\end{equation}

To derive the second Maxwell field equation we will make use of (\ref{M1.8}%
), (\ref{M1.1}) and (\ref{M1.5}):%
\begin{eqnarray}
\nabla \times B &=&\nabla \times (\nabla \times A)=\nabla <\nabla ,A>-\nabla
^{2}A=  \notag \\
&=&\nabla (-\frac{1}{c}\frac{\partial W}{\partial t})-\nabla ^{2}A=\frac{1}{c%
}\frac{\partial }{\partial t}(-\nabla W-\frac{1}{c}\frac{\partial A}{%
\partial t}+\frac{1}{c}\frac{\partial A}{\partial t})-\nabla ^{2}A=  \notag
\\
&=&\frac{1}{c}\frac{\partial E}{\partial t}+(\frac{1}{c^{2}}\frac{\partial
^{2}A}{\partial t^{2}}-\nabla ^{2}A).  \label{M1.9}
\end{eqnarray}%
We have from (\ref{M1.5}), (\ref{M1.6}) and (\ref{M1.3}) that
\begin{equation*}
<\nabla ,\frac{1}{c}\frac{\partial E}{\partial t}>=\frac{1}{c}\frac{\partial
\rho }{\partial t}=-\frac{1}{c}<\nabla ,J>,
\end{equation*}%
or
\begin{equation}
<\nabla ,-\frac{1}{c^{2}}\frac{\partial ^{2}A}{\partial t^{2}}-\nabla (\frac{%
1}{c}\frac{\partial W}{\partial t})+\frac{1}{c}J>=0.  \label{M1.10}
\end{equation}%
Now making  use of (\ref{M1.1}), from (\ref{M1.10}) we obtain that%
\begin{eqnarray}
&<&\nabla ,-\frac{1}{c^{2}}\frac{\partial ^{2}A}{\partial t^{2}}-\nabla (%
\frac{1}{c}\frac{\partial W}{\partial t})+\frac{1}{c}J>=<\nabla ,-\frac{1}{%
c^{2}}\frac{\partial ^{2}A}{\partial t^{2}}+\nabla <\nabla ,A>+\frac{1}{c}J=
\notag \\
&=&<\nabla ,-\frac{1}{c^{2}}\frac{\partial ^{2}A}{\partial t^{2}}+\nabla
^{2}A+\nabla \times (\nabla \times A)+\frac{1}{c}J>=  \notag \\
&=&<\nabla ,-\frac{1}{c^{2}}\frac{\partial ^{2}A}{\partial t^{2}}+\nabla
^{2}A+\frac{1}{c}J>=0.  \label{M1.11}
\end{eqnarray}%
Thereby, equation (\ref{M1.11}) yields%
\begin{equation}
\frac{1}{c^{2}}\frac{\partial ^{2}A}{\partial t^{2}}-\nabla ^{2}A=\frac{1}{c}%
(J+\nabla \times S)  \label{M1.12}
\end{equation}%
for some smooth vector function $S:\mathbb{R}^{3}\times \mathbb{R}%
\rightarrow \mathbb{R}^{3}.$ Here we need to note that continuity equation (%
\ref{M1.3}) is defined, concerning the current density vector $J:\mathbb{R}%
^{3}\times \mathbb{R}\rightarrow \mathbb{R}^{3},$ up to a vorticity
expression, that is $J\simeq J+\nabla \times S$ and equation (\ref{M1.12})
can finally be rewritten down as
\begin{equation}
\frac{1}{c^{2}}\frac{\partial ^{2}A}{\partial t^{2}}-\nabla ^{2}A=\frac{1}{c}%
J.  \label{M1.13}
\end{equation}%
Having substituted  (\ref{M1.13}) into (\ref{M1.9}) we obtain the second
Maxwell field equation

\begin{equation}
\nabla \times B-\frac{1}{c}\frac{\partial E}{\partial t}=\frac{1}{c}J.
\label{M1.14}
\end{equation}%
In addition, from (\ref{M1.8}) one also finds the magnetic no-charge
relationship%
\begin{equation}
<\nabla ,B>=0.  \label{M1.15}
\end{equation}

Thus, we have derived all the Maxwell electromagnetic field equations from
our three main principles (\ref{M1.1}), (\ref{M1.2}) and (\ref{M1.3}). The
success of our undertaking will be more impressive if we adapt our results
to those following from the well known relativity theory in the case of
point charges or masses. Below we will try to demonstrate the corresponding
derivations based on some completely new physical conceptions of the vacuum
medium first discussed in \cite{Re,PB}.

\begin{remark}
It is interesting to analyze a partial case of the first field theory vacuum
principle (\ref{M1.1}) when the following local conservation law for the
scalar potential field function $W:\mathbb{R}^{3}\times \mathbb{R}%
\rightarrow \mathbb{R}$ holds:%
\begin{equation}
\frac{d}{dt}\int_{\Omega _{t}}Wd^{3}r=0,  \label{M1.16}
\end{equation}%
where $\Omega _{t}\subset \mathbb{R}^{3}$ is any open domain in space $%
\mathbb{R}^{3}$ with the smooth boundary $\partial \Omega _{t}$ \ for all $%
t\in \mathbb{R}$ and $d^{3}r$ \ is the standard volume measure in $\mathbb{R}%
^{3}$ in a vicinity of the point $r\in \Omega _{t}.$
\end{remark}

Having calculated expression (\ref{M1.16}) we obtain the following
equivalent continuity equation
\begin{equation}
\frac{1}{c}\frac{\partial W}{\partial t}+<\nabla ,\frac{v}{c}W>=0,
\label{M1.17}
\end{equation}%
where $\nabla :=\nabla _{r}$ is, as above, the gradient operator and $%
v:=dr/dt$ is the velocity vector of a vacuum medium perturbation at point $%
r\in \mathbb{R}^{3}$ carrying the field potential quantity $W.$ Comparing
now equations (\ref{M1.1}), (\ref{M1.17}) and using equation (\ref{M1.3}) we
can make the suitable identifications:
\begin{equation}
A=\frac{v}{c}W,\text{ \ \ \ \ }J=\rho v,  \label{M1.18}
\end{equation}%
well known from the classical electrodynamics and superconductivity theory
\cite{Fe1,Ge}. Thus, we are  faced with a new physical interpretation of the
conservative electromagnetic field theory when the vector \ potential $A:%
\mathbb{R}^{3}\times \mathbb{R}\rightarrow \mathbb{R}^{3}$ is completely
determined via expression (\ref{M1.18}) by the scalar field potential
function $W:\mathbb{R}^{3}\times \mathbb{R}\rightarrow \mathbb{R}.$ It is
also evident that all  the Maxwell electromagnetism filed equations derived
above hold as well  in the case (\ref{M1.18}), as it was first demonstrated
in \cite{Re} (but with some mathematical inaccuracies) and in \cite{PB}.

Consider now the conservation equation (\ref{M1.16}) jointly with the
related integral "vacuum momentum" conservation relationship
\begin{equation}
\frac{d}{dt}\int_{\Omega _{t}}(\frac{vW}{c^{2}})d^{3}r=0,~~~\Omega
_{t}|_{t=0}=\Omega _{0},  \label{M1.19}
\end{equation}%
where, as above,  $\Omega _{t}\subset \mathbb{R}^{3}$ is for any time $t\in
\mathbb{R}$ an open domain with the smooth boundary $\partial {\Omega _{t}},$
whose evolution is governed by the equation
\begin{equation}
dr/dt=v(r,t)  \label{M1.20}
\end{equation}%
for all $x\in \Omega _{t}$ and $t\in \mathbb{R},$ as well as by the initial
state of the boundary $\partial {\Omega }_{0}.$ As a result of relation  (%
\ref{M1.19}) one obtains the new continuity equation
\begin{equation}
\frac{d(vW)}{dt}+vW<\nabla ,v>=0.  \label{M1.21}
\end{equation}%
Now making  use of (\ref{M1.17}) in the equivalent form
\begin{equation*}
\frac{dW}{dt}+W<\nabla ,v>=0,
\end{equation*}%
we  finally obtain a very interesting local conservation relationship \
\begin{equation}
dv/dt=0\   \label{M1.22}
\end{equation}%
on the vacuum matter perturbations velocity $v=dr/dt,$ which holds  for all
values of the time parameter $t\in \mathbb{R}.$ As it is easy to observe,
the obtained relationship completely coincides with the well-known
hydrodynamic equation \cite{MC} of ideal compressible liquid without any
external exertion, that is, any external forces and field "pressure" \ are
equally identical to zero. We received a natural enough result where the
propagation velocity of the vacuum field matter is constant and equals
exactly $v=c,$ that is the light velocity in the vacuum,  if to recall the
starting wave equation (\ref{M1.2}) owing to which the small vacuum field
matter perturbations propagate in the space with the light velocity.

\section{Special relativity theory and dynamical field equations}

From classical electrodynamics we know that the main dynamical relationship
relates the particle mass acceleration to the Lorentz force which strongly
depends on the absolute \bigskip charge velocity. For the electrodynamics to
be independent on the reference system physicists were forced to reject the
Galilean transformations and replace them with the artificially postulated
Lorentz transformations. This resulted later in the Einstein relativity
theory which has partly reconciled the problems concerned with deriving true
dynamical equations for a charged point particle.

We will now start from the scalar field vacuum medium function $W:\mathbb{R}%
^{3}\times \mathbb{R}\rightarrow \mathbb{R}$ in the conservation condition
case (\ref{M1.16}) discussed above. This means, obviously, that the vacuum
medium field vector potential $A:\mathbb{R}^{3}\times \mathbb{R}\rightarrow
\mathbb{R}^{3},$ charge and current densities $(\rho ,J):\mathbb{R}%
^{3}\times \mathbb{R}\rightarrow \mathbb{R}$ $\times \mathbb{R}^{3}$ are
related owing to expressions (\ref{M1.18}).

Consider now  vacuum field medium conservation equations (\ref{M1.17}) and (%
\ref{M1.2}) at the density $\rho =0:$%
\begin{eqnarray}
-\frac{1}{c^{2}}\frac{\partial ^{2}W}{\partial t^{2}} &=&\frac{1}{c^{2}}%
\frac{\partial }{\partial t}(-\frac{\partial W}{\partial t})=\frac{1}{c^{2}}%
\frac{\partial }{\partial t}(<\nabla ,vW>)=  \notag \\
&=&<\nabla ,\frac{\partial }{\partial t}(\frac{Wv}{c^{2}})>=-<\nabla ,\nabla
W>.  \label{M2.1}
\end{eqnarray}%
From relation (\ref{M2.1}) it follows that%
\begin{equation}
\frac{\partial }{\partial t}(\frac{Wv}{c^{2}})+\nabla W=\nabla \times F,
\label{M2.2}
\end{equation}%
where $F:\mathbb{R}^{3}\times \mathbb{R}\rightarrow \mathbb{R}^{3}$ is some
smooth function, which we put, by definition, to be zero owing to the
\textit{a priori} assumed vortexless vacuum medium dynamics. So, our
dynamical equation on the vacuum medium scalar field function $W:\mathbb{R}%
^{3}\times \mathbb{R}\rightarrow \mathbb{R}$ looks like
\begin{equation}
\frac{\partial }{\partial t}(\frac{Wv}{c^{2}})+\nabla W=0.  \label{M2.3}
\end{equation}

Consider now a charged point particle $q$ in the space point $%
r=R(t):=R_{0}+\dint\limits_{0}^{t}u(t)dt\in \mathbb{R}^{3},$ depending on
time parameter $t\in \mathbb{R}$ and initial point $R_{0}\in \mathbb{R}^{3}$
at\textbf{\ }$t=0.$ \ Since the vacuum medium field is described by means of
the potential field function $W:\mathbb{R}^{3}\times \mathbb{R}\rightarrow
\mathbb{R},$ which is naturally disturbed by the charged particle $q,$ we
will model this fact approximately as the following resulting functional
relationship:%
\begin{equation}
W(r,t)=\tilde{W}(r,R(t))  \label{M2.4}
\end{equation}%
for some scalar function $\tilde{W}:\mathbb{R}^{3}\times \mathbb{R}%
^{3}\rightarrow \mathbb{R}.$ This function must satisfy equation (\ref{M1.17}%
), that is
\begin{equation}
<\frac{\partial \tilde{W}}{\partial R},u>+<\nabla ,\tilde{W}v>=0.
\label{M2.5}
\end{equation}%
As we are interested in differential properties of the function $\tilde{W}:%
\mathbb{R}^{3}\times \mathbb{R}^{3}\rightarrow \mathbb{R}$ as $r\rightarrow
R(t)\in \mathbb{R}^{3},$ where the charged point particle is located, we
obtain from (\ref{M2.5}) that
\begin{equation*}
<\frac{\partial \tilde{W}}{\partial R}+\frac{\partial \tilde{W}}{\partial r}%
,u>|_{r\rightarrow R(t)}+\tilde{W}<\nabla ,v>|_{r\rightarrow R(t)}=0,
\end{equation*}%
giving rise to the relationship
\begin{equation}
\frac{\partial \tilde{W}}{\partial R}=-\frac{\partial \tilde{W}}{\partial r}
\label{M2.6}
\end{equation}%
as $r\rightarrow R(t),$ since $v|_{r\rightarrow R(t)}\rightarrow
dR(t)/dt:=u(t)$ and $<\nabla ,v>|_{r\rightarrow R(t)}\rightarrow <\nabla
,u(t)>=0$ for all $t\in \mathbb{R}.$

Returning now to equation (\ref{M2.3}) we can write,  owing to (\ref{M2.6}),
that
\begin{equation}
\begin{array}{c}
\frac{1}{c^{2}}\left. \left( \frac{\partial \tilde{W}}{\partial t}v+\tilde{W}%
\frac{\partial v}{\partial t}\right) \right\vert _{r\text{ }\rightarrow
R(t)}=\frac{1}{c^{2}}\left. \left( -<\frac{\partial \tilde{W}}{\partial r}%
,v>v+\tilde{W}\frac{\partial v}{\partial t}\right) \right\vert
_{r\rightarrow R(t)}= \\
=\frac{1}{c^{2}}\left. \left( <\frac{\partial \tilde{W}}{\partial R},u>u+%
\tilde{W}\frac{du}{dt}\right) \right\vert _{r\rightarrow R(t)}\Rightarrow
\frac{1}{c^{2}}\frac{d}{dt}(\bar{W}u)=-\left. \frac{\partial \tilde{W}}{%
\partial r}\right\vert _{r\rightarrow R(t)}=\frac{\partial \bar{W}}{\partial
R},%
\end{array}
\label{M2.7}
\end{equation}%
where we put, by definition, $\bar{W}:=\tilde{W}(r,R(t))|_{r\rightarrow
R_{0}}.$ Thus, we obtained from (\ref{M2.7}) that the function $\bar{W}:%
\mathbb{R}^{3}\rightarrow \mathbb{R}$ satisfies the determining dynamical
equation
\begin{equation}
\frac{d}{dt}(-\frac{\bar{W}}{c^{2}}u)=-\frac{\partial \bar{W}}{\partial R}
\label{M2.8}
\end{equation}%
at point $R(t)\in \mathbb{R}^{3},$ $t\in \mathbb{R},$ of the point charge $q$
location.

Now we need to proceed with our calculations \ and would like to make the
following very important \textbf{assumption:} we will interpret the quantity
$-\frac{\bar{W}}{c^{2}}$ as the real "dynamical" mass of our point charge $q$
at point $R(t)\in \mathbb{R}^{3},$ that is
\begin{equation}
m:=-\frac{\bar{W}}{c^{2}}.  \label{M2.9}
\end{equation}%
This, in part, means that the whole observed particle mass $m$ depends only
on the vacuum field potential $\bar{W}$ owing to both  its gravitational and
electromagnetic interactions with long distant and closely ambient it
material particles! This statement, evidently, in many points coincides with
the well-known Mach principle \cite{Br,Re,PB} and formalizes it concerning
the real field structure of vacuum. We need here to mention that this idea
was also earlier claimed, but not realized practically, by L. Brillouin in
\cite{Br}. We press here that no assumption about the equivalence of the
inertial mass and gravitational mass is made and, moreover, such a kind of
statement  is completely \ alien within the theory devised here and in \cite%
{Re,PB}.

Using further (\ref{M2.9}) we can rewrite equation (\ref{M2.8}) as
\begin{equation}
\frac{dp}{dt}=-\frac{\partial \bar{W}}{\partial R},  \label{M2.10}
\end{equation}%
where the quantity $p:=mu$ has the natural momentum interpretation.

The obtained equation (\ref{M2.10}) is very interesting from the dynamical
point of view. Really, from equation (\ref{M2.10}) we obtain that
\begin{equation}
<u,\frac{d}{dt}(mu)>=c^{2}<\frac{\partial m}{\partial R},u>=c^{2}\frac{dm}{dt%
}.  \label{M2.11}
\end{equation}%
As a result of (\ref{M2.11}) we easily derive, following \cite{Re,BP}, the
conservative relationship%
\begin{equation}
\frac{d}{dt}\left( m\sqrt{1-\frac{u^{2}}{c^{2}}}\right) =0  \label{M2.12}
\end{equation}%
for all $t\in \mathbb{R}.$ Really, based on (\ref{M2.11}), we have that%
\begin{equation}
m<u,\frac{du}{dt}>+<u,u>\frac{dm}{dt}=c^{2}\frac{dm}{dt},  \label{M2.12a}
\end{equation}%
\ or, equivalently,
\begin{equation}
\frac{1}{2}m\frac{du^{2}}{dt}-(c^{2}-u^{2})\frac{dm}{dt}=0.  \label{M2.12b}
\end{equation}%
As a result of (\ref{M2.12b}), we obtain
\begin{equation}
-\frac{1}{2}\frac{1}{(1-\frac{u^{2}}{c^{2}})}\frac{d}{dt}(\frac{u^{2}}{c^{2}}%
)+\frac{1}{m}\frac{dm}{dt}=0,  \label{M2.12c}
\end{equation}%
giving rise, by means of simple integrating, to the following differential
expression:%
\begin{equation}
\frac{d}{dt}\ln (\sqrt{1-\frac{u^{2}}{c^{2}}})+\frac{d\ln m}{dt}=\frac{d}{dt}%
\ln \left( m\sqrt{1-\frac{u^{2}}{c^{2}}}\right) =0.  \label{M2.12d}
\end{equation}%
The latter is, evidently, equivalent to result (\ref{M2.12}), that is the
quantity
\begin{equation}
m\sqrt{1-\frac{u^{2}}{c^{2}}}=m_{0}  \label{M2.13}
\end{equation}%
is constant for all $t\in \mathbb{R},$ giving rise to the well known
relativistic expression for the mass of a point particle:%
\begin{equation}
m=\frac{m_{0}}{\sqrt{1-\frac{u^{2}}{c^{2}}}}.  \label{M2.14}
\end{equation}%
As we can see, the point particle mass $m$ depends, in reality, not on the
coordinate $R(t)\in \mathbb{R}^{3}$ of the point particle $q,$ but on its
velocity $u:=dR(t)/dt.$ Since the field potential $\bar{W}:\mathbb{R}%
^{3}\rightarrow \mathbb{R}$ \ consists of two parts
\begin{equation}
\bar{W}=\bar{W}_{0}+\Delta \bar{w},  \label{M2.15}
\end{equation}%
where $\bar{W}_{0}:\mathbb{R}^{3}\rightarrow \mathbb{R}$ is constant and
responsible for the external influence of all long distant objects in the
Universe upon the point particle $q$ and $\Delta \bar{w}:\mathbb{R}%
^{3}\rightarrow \mathbb{R}$ is responsible for the local field potential
perturbation by the point charge $q$ and its closest ambient \ neighborhood.
Then, obviously,
\begin{equation}
\Delta m:=m-m_{0}=-\Delta \bar{w}/c^{2}  \label{M2.16}
\end{equation}%
is the strictly dynamical mass component belonging to the point particle $q.$
Moreover, since the full momentum $p=mu$ satisfies equation (\ref{M2.10}),
one can easily obtain that the quantity
\begin{equation}
\bar{W}^{2}-c^{2}p^{2}=E_{0}^{2}  \label{M2.17}
\end{equation}%
is not depending on time $t\in \mathbb{R},$ that is, $dE_{0}/dt=0,$ where $%
E_{0}:=m_{0}c^{2}$ is the so-called\ \cite{Fe1,Di,Fo1,TW} "\textit{internal
\ energy"} of a point particle $q.$ The result (\ref{M2.17}) demonstrates
the important property of the energy essence: the point particle $q$ is, in
reality, endowed with the only dynamical energy $\Delta E:=\Delta mc^{2}.$
Concerning this  "internal" particle energy $E_{0}=m_{0}c^{2}$ we see that
it has nothing to do with the real particle energy, since its origin is
determined completely owing to the long distant objects of the Universe and
can not be used for any physical processes, contrary to the known Einstein
theory statements about a "huge" internal energy stored inside the particle
mass. Equivalently, the Einstein theory statement about the "equivalence" of
the mass and the "internal" energy of particle appears to be senseless,
since the main part of the field potential function $\bar{W}:\mathbb{R}%
^{3}\rightarrow \mathbb{R}$ at the location point of the\ point particle $q$
is constant and results owing to the long distant objects in the Universe,
which obviously can not be used for so called "practical applications".

Nonetheless, we \ have observed above, as a by-product, the well known
"relativistic" effect of the particle mass growth depending on the particle
velocity in the form (\ref{M2.14}). As it was already mentioned in \cite%
{Re,PB} this "mass growth" is, in reality, completely of dynamical nature
and is not a consequence of the Lorentz transformations, as it was stated
within the Einstein SRT. Moreover, we can state that all of the so-called
"relativistic" effects have  nothing to do with  the mentioned above Lorentz
transformations and with such artificial "effects" as length "shortening"
and time "slowing". There is also no reasonable cause to identify the
particle mass with its real energy and vice versa. Concerning the
interesting physical effect called particles "annihilation" we need  to
stress\ here that it has also nothing to do with the transformation of
particles masses into energy. The field theoretical explanation of this
phenomenon consists in creating their very special bonding state, whose
interaction with ambient objects is vanishing. As a result the visible
inertial or dynamical mass of this bound state is also zero, exactly what
the experiment shows, and nothing else. Inversely, if an intensive enough
photon meets such a bound state of two particles, it can break them back
into two separate particles, what the experiment  shows to happen. Here we
recall a similar analogy borrowed from the modern quantum physics of
infinite particle systems described by means of the\ second quantization
scenario \cite{Fo,Be,BP,BPT} suggested in 1932 by V. Fock. Within this
scenario there are also realized creating-annihilation effects which are
present owing to the inter-particle interaction forces. Moreover, as we know
from the modern  superconductivity theory within this picture one can
describe special bound states of particles, so-called "Cooper pairs", whose
interaction with  each other completely vanishes and whose combined mass
strongly\ differs from the sum of the suitable components and equals the
so-called "effective" compound mass, depending strongly on the potential
field intensity inside the superconductor matter.

We now proceed to discuss  the  relation (\ref{M2.17})\ derived above, where
the conserved quantity $E_{0}$ is naturally interpreted as the total energy
of a particle moving with velocity $u:=$ $dR(t)/dt$ in vacuum endowed with
the field potential $\bar{W}$ in a vicinity of the particle $q$ located at
point $R(t)\in \mathbb{R}^{3}.$ We see that the total particle energy $E_{0}$
strictly depends  on both the field potential $\bar{W}$ and  its velocity $u,
$ as the particle momentum $p=m_{0}u/\sqrt{1-\frac{u^{2}}{c^{2}}}$ depends
strictly relativistically on its velocity $u.$

As it was \ mentioned by L. Brillouin in \cite{Br} the Einstein SRT
postulates that the total energy $E_{0}$ of a moving particle in \ a
potential field $U$ equals $E_{0}=m_{0}c^{2}/\sqrt{1-\frac{u^{2}}{c^{2}}}+U.$
He writes: \textit{"This means that any possibility of existing the particle
mass related with the external potential energy is completely excluded...
Thereby, the strange situation appears: the internal (particle) energy is
endowed with the mass but the external - is not". }(The citation is taken
from\textit{\ }\cite{Br}).

Contrary to this inference from the Einstein SRT, \ the relation (\ref{M2.17}%
) obtained above \ naturally takes into account both the kinetic energy of
the particle motion with velocity $u$ and the field potential energy $\bar{W}
$ in a vicinity of the particle $q$ located at point $R(t)\in \mathbb{R}^{3}.
$

Moreover, L. Brillouin in \cite{Br} writes: "\textit{...Einstein tends by
any way to reduce gravity to geometry by means of changing the Newtonian
gravitational potential by a tensor potential of second order, realizing the
joint description of gravity and geometry; this is achieved owing to the
appearance of a huge gap between gravity and electromagnetism.... The
(Einstein) article - a genuine mathematical work but its application to the
physical reality \ remains to be open."}\textbf{\ } \ Later V. Fock in his
famous book \cite{Fo1} tried to rescue the situation that  arisen with
gravity and electromagnetism but his approach was also based on the Einstein
geometrization ideas and no \ real success was achieved.  Evaluating the
gravity and electromagnetism theories state of art L. Brillouin in \cite{Br}
states  that "\textit{In general, the necessity of considering a curved
space-time Universe is still not proved; the physical meaning of the general
relativity is still very vague". }

Having analyzed, from this point of view,  the results formulated above we
see that all of the Einstein SRT statements were obtained within the
classical Euclidean space-time scenario and no four-dimensional space-time
geometry, like the invariance of the four-dimensional metric interval with
respect to the Lorentz transformations, was  involved. \ We will show below
that within the approaches devised above and in \cite{Re,PB} we will derive
the next very important result of the SRT and Maxwell electromagnetism
theory, namely that related to the nature of the Lorentz force acting on a
moving in space charged particle. As is well known \cite{Br,Fo1,BD,Re},
wishing to make the Lorentz force expression compatible with the postulated
relativity principles appeared to be decisive in Einstein's endeavors to
construct his SRT and later the general relativity theory.

As a very interesting aspect of the vacuum field theory devised above and in
\cite{PB,Re} we need here to mention a very close relationship of
electromagnetic and gravity fields. Namely, the vacuum field potential, in
general, consists of two components: $W:=W_{g}+W_{em},$ where $W_{g}$
corresponds to the gravity interaction between material particles and $W_{em}
$ corresponds to the electromagnetic interaction between their charges,
where we have accepted that these two physical vacuum realities are
different and independent. Thecorresponding full mass of a particle is given
then by expression (\ref{M2.9}) in the form%
\begin{equation}
m:=-\bar{W}/c^{2}=-(\bar{W}_{g}+\bar{W}_{em})/c^{2},  \label{M2.18}
\end{equation}%
including both the gravity and electromagnetic interactions between
particles. Then, following the derivation of electromagnetic Maxwell
equations above, based on the electromagnetic vacuum field potential $%
W:=W_{em},$ we can derive, in  the same way,  the corresponding Maxwell type
gravimagnetic equations on the gravity potential $W_{g},$  as follows%
\begin{equation}
\begin{array}{c}
\frac{1}{c^{2}}\frac{\partial ^{2}W_{g}}{\partial t^{2}}-\nabla
^{2}W_{g}=\rho _{g},\frac{1}{c^{2}}\frac{\partial ^{2}A_{g}}{\partial t^{2}}%
-\nabla ^{2}A_{g}=J_{g}, \\
\text{ } \\
\nabla \times B_{g}-\frac{1}{c}\frac{\partial E_{g}}{\partial t}=\frac{1}{c}%
J_{g},\frac{1}{c}\frac{\partial B_{g}}{\partial t}-\nabla \times E_{g}=0, \\
\\
<\nabla ,E_{g}>=\rho _{g},\text{ \ }<\nabla ,B_{g}>=0,%
\end{array}
\label{M2.19}
\end{equation}%
where, by definition,
\begin{equation}
E_{g}:=-\nabla W_{g}-\frac{1}{c}\frac{\partial A_{g}}{\partial t},\text{ }%
B_{g}:=\nabla \times A_{g}.  \label{M2.20}
\end{equation}%
Here $\rho _{g}$ and $J_{g}$ denote the gravity particle mass density and
mass current density, \ respectively. The form of governing equations (\ref%
{M2.19}) is, evidently, strongly motivated by the well known similarity
between the Coulomb electrical and Newtonian gravitational forces
expressions, and was  very deeply previously discussed in \cite{Br,Ca,Re,Pe}.

As already shown above, in the conservative case, the following
representations
\begin{equation}
J_{g}=\rho _{g}v,\text{ \ }A_{g}=\frac{v}{c}W_{g},  \label{M2.21}
\end{equation}%
similar to (\ref{M1.18}), hold. As a consequence, we can state that the
gravitational waves exist, propagating in vacuum with the same light
velocity $c$ \ as electromagnetic waves, and are described by means of the
Maxwell type gravimagnetic equations (\ref{M2.19}). Here we  note that
similar inferences have been done many years ago  in classical oeuvres of
famous physicists of the XIX century J.C. Maxwell \cite{Max} and O.
Heaviside \cite{He}, as well as in works of other physicists of the past and
present centuries \cite{BL,Pe,Bu,Re}.

\bigskip

\section{The Lorentz force and the relativity theory principles revisited}

It is a well known fact that the Einstein special relativity theory is
applicable only for physical processes related to each other by means of the
inertial reference systems, moving with constant velocities. In this case
one can make use of the Lorentz transformations and calculate the components
of suitable four-vectors and the resulting mass growth of particles owing to
formula (\ref{M2.11}). A nontrivial problem arises when we wish to analyze
these quantities with respect to non-inertial reference systems moving with
some nonzero acceleration. Below we will revisit this problem from the
vacuum field theory scenario devised above and show that the whole "special"
relativity theory emerges as its partial case or by-product and is free of
the artificial "inertial reference systems" problems mentioned above.

Really, our\ vacuum field theory \ structure is described by the dynamical
equation (\ref{M2.3}), which we would like to investigate in a neighborhood
of two interacting to each other point particles $q_{f}$ at point $%
R_{f}(t)\in \mathbb{R}^{3}$ and $\ q$ at point $R(t):=R_{0}+\dint%
\limits_{0}^{t}u(t)dt\in \mathbb{R}^{3},$ respectively, depending on time
parameter $t\in \mathbb{R}$ and initial point $R_{0}\in \mathbb{R}^{3}$ at%
\textbf{\ }$t=0.$ As was already done in Section 2 we assume that the\
vacuum potential field function $W:\mathbb{R}^{3}\times \mathbb{R}%
\rightarrow \mathbb{R}$ can be represented as $W=\tilde{W}(r;R_{f}(t),R(t))$
for some function $\tilde{W}:\mathbb{R}^{3}\times \mathbb{R}^{3}\times
\mathbb{R}^{3}\rightarrow \mathbb{R}$ and all $t\in \mathbb{R}.$ Then, based
on the continuity equation (\ref{M1.17}) we obtain%
\begin{equation}
<\frac{\partial \tilde{W}}{\partial R_{f}},u_{f}>+<\frac{\partial \tilde{W}}{%
\partial R},u>+<\frac{\partial \tilde{W}}{\partial r},v>+\tilde{W}<\nabla
,v>=0.  \label{M3.1}
\end{equation}%
We will now be interested in the potential field function $\tilde{W}:\mathbb{%
R}^{3}\times \mathbb{R}^{3}\times \mathbb{R}^{3}\rightarrow \mathbb{R}$ in a
vicinity of the relative distance vector $\tilde{R}(t):=R(t)-R_{f}(t)\in
\mathbb{R}^{3},$ keeping in mind that the interaction between particles $%
q_{f}$ and $q_{\text{ }}$depends on this relative interparticle distance $%
\tilde{R}(t)\in \mathbb{R}^{3}.$ From (\ref{M3.1}), as \  vector $%
r\rightarrow \tilde{R}(t),$ we easily derive that
\begin{equation}
\left. \frac{\partial \tilde{W}}{\partial R_{f}}+\frac{\partial \tilde{W}}{%
\partial R}\right\vert _{r\rightarrow \tilde{R}(t)}=0,\text{ \ \ \ \ }\left.
\frac{\partial \tilde{W}}{\partial R}+\frac{\partial \tilde{W}}{\partial r}%
\right\vert _{r\rightarrow \tilde{R}(t)}=0.  \label{M3.2}
\end{equation}%
Combining relations (\ref{M3.2}) with the dynamical field equations (\ref%
{M2.3}) we obtain that
\begin{equation*}
\begin{array}{c}
\frac{1}{c^{2}}\frac{\partial }{\partial t}\left. \left( \tilde{W}v\right)
\right\vert _{r\text{ }\rightarrow \tilde{R}(t)}=\frac{1}{c^{2}}\left.
\left( <\frac{\partial \tilde{W}}{\partial R_{f}},u_{f}>v+<\frac{\partial
\tilde{W}}{\partial R},u>v+\tilde{W}\frac{\partial v}{\partial t}\right)
\right\vert _{r\rightarrow \tilde{R}(t)}= \\
=\frac{1}{c^{2}}\left. \left( <\frac{\partial \tilde{W}}{\partial \tilde{R}}%
,u-u_{f}>(u-u_{f})\right) \right\vert _{r\rightarrow \tilde{R}(t)}=-\left.
\frac{\partial \tilde{W}}{\partial r}\right\vert _{r\rightarrow \tilde{R}%
(t)}=\left. \frac{\partial \tilde{W}}{\partial R}\right\vert _{r\rightarrow
\tilde{R}(t)},%
\end{array}%
\end{equation*}%
whence one derives the new dynamical equation
\begin{equation}
\frac{d}{dt}(-\frac{\bar{W}}{c^{2}}(u-u_{f}))=-\frac{\partial \bar{W}}{%
\partial R}  \label{M3.3}
\end{equation}%
on the resulting function $\bar{W}:=\tilde{W}|_{r\rightarrow \tilde{R}(t)}$

Equation (\ref{M3.3}) possesses a very important feature of depending on the
only relative quantities not depending on the reference system. Moreover, we
have not, on the whole,  met the necessity to use other transformations of
coordinates different from the Galilean transformations. We mention here
that the  dynamical equation (\ref{M3.3}) was also derived in \cite{Re}
making use of some not completely true relationships and mathematical
manipulations. But the main corollary of \cite{Re} and our derivation \cite%
{PB}, saying that equation (\ref{M3.3}) fits for all reference systems, both
inertial and accelerated, appears to be fundamental and gives rise to new
unexpected results in the modern electrodynamics and gravity theory. Below
we will proceed with one of very important relativity physics aspects,
concerned with the well-known Lorentz force expression measuring the action
exerted by external electromagnetic field on a charged point particle $q$ at
space point $R(t)\in \mathbb{R}^{3}$ for any time moment $t\in \mathbb{R}.$

To do this we accept, owing to the vacuum field theory, that the resulting
potential field function $\bar{W}:\mathbb{R}^{3}\rightarrow \mathbb{R}$ can
be represented in the vicinity of the charged point particle $q$ as
\begin{equation}
\bar{W}=\bar{W}_{0}+q\varphi ,  \label{M3.3a}
\end{equation}%
where $\varphi :\mathbb{R}^{3}\rightarrow \mathbb{R}$ is a suitable local
electromagnetic field potential and $\bar{W}_{0}:\mathbb{R}^{3}\rightarrow
\mathbb{R}$ is a constant vacuum field potential owing to the particle
interaction with the external distant Universe objects. Then, having
substituted (\ref{M3.3a}) into \ the main dynamical field equation (\ref%
{M3.3}) we obtain that
\begin{eqnarray}
\frac{d}{dt}(-\frac{\bar{W}}{c^{2}}u) &=&\frac{d}{dt}(-\frac{\bar{W}}{c^{2}}%
u_{f})-\nabla \bar{W}=-\nabla \bar{W}+\frac{\partial }{\partial t}(-\frac{%
\bar{W}}{c^{2}}u_{f})+<u,\nabla >(-\frac{\bar{W}}{c^{2}}u_{f})=  \notag \\
&=&-\nabla \bar{W}+\frac{1}{c}\frac{\partial }{\partial t}(-\frac{\bar{W}}{c}%
u_{f})-u\times (u_{f}\times \nabla \frac{\bar{W}}{c^{2}})-<u,u_{f}>\nabla
\bar{W}=  \notag \\
&=&-\nabla \bar{W}(1+\frac{<u,u_{f}>}{c^{2}})+\frac{1}{c}\frac{\partial }{%
\partial t}(-\frac{\bar{W}}{c}u_{f})-\frac{1}{c^{2}}u\times (u_{f}\times
\nabla \bar{W})=  \notag \\
&=&-q\nabla \varphi (1+\frac{<u,u_{f}>}{c^{2}})-\frac{q}{c}\frac{\partial }{%
\partial t}(\frac{\varphi }{c}u_{f})+\frac{q}{c}u\times (\nabla \times \frac{%
\varphi u_{f}}{c})=  \notag \\
&=&-q\nabla \varphi (1+\frac{<u,u_{f}>}{c^{2}})-\frac{q}{c}\frac{\partial A}{%
\partial t}+\frac{q}{c}u\times (\nabla \times A),  \label{M3.3b}
\end{eqnarray}%
where we denoted $u:=dR(t)/dt,$ \ $u_{f}:=dR_{f}(t)/dt,$ $\nabla :=\partial
/\partial R=-\partial /\partial R_{f}$ and $A:=\varphi u_{f}/c$  to be  the
related magnetic potential. Since we have already shown that the Lorentz
force
\begin{equation*}
F:=\frac{d}{dt}(-\frac{\bar{W}}{c^{2}}u)=\frac{d}{dt}\left( \frac{m_{0}u}{%
\sqrt{1-\frac{u^{2}}{c^{2}}}}\right)
\end{equation*}%
is given by expression (\ref{M3.3b}), it can be rewritten  in the form
\begin{eqnarray}
F &=&\frac{d}{dt}\left( \frac{m_{0}u}{\sqrt{1-\frac{u^{2}}{c^{2}}}}\right)
=qE+\frac{q}{c}u\times B-\frac{q}{c^{2}}\nabla \varphi <u,u_{f}>=
\label{M3.4} \\
&=&qE+\frac{q}{c}u\times B-\frac{q}{c}\nabla <u,A>,  \notag
\end{eqnarray}%
which was derived also in \cite{Re} and where we put, by definition, $%
E:=-\nabla \varphi -\frac{1}{c}\frac{\partial A}{\partial t},$ \ $B:=\nabla
\times A,$ being respectively the suitable electric and magnetic vector
fields.

The resulting expression (\ref{M3.4}) is almost completely equivalent to the
well-known \cite{AB} classical Lorentz force expression $F$ up to the
additional "inertial" term
\begin{equation}
F_{c}:=-\frac{q}{c}\nabla <u,A>,  \label{M3.5}
\end{equation}%
which is absent in the Einstein  special relativistic theory. Namely, owing
to the absence of the  term (\ref{M3.5}) the classical relativistic Lorentz
force expression in the four-vector form was not invariant with respect to
any reference frame transformations, except inertial ones. And, as  was
noticed in \cite{Re,Br}, owing only to this crucial fact   A. Einstein
introduced the Lorentz transformations and related with them visible length
shortening and time slowing effects!  Moreover, they gave rise to such
strange enough and non-adequate notions as non-Euclidean time-spaces \cite%
{Ho,Me,Ba1,Br}, black holes \cite{Da,Gr,Ho,Ba} and some other singular and
nonphysical objects.

Now based  on the Lorentz force expression (\ref{M3.4}) we can easily obtain
the corresponding energy conservation law for a moving charged point
particle $q$ and the ambient magnetic field:%
\begin{equation}
\frac{dE_{0}^{2}}{dt}=-2c<\nabla _{R_{f}}\bar{W},qA>=-\frac{dE_{f}^{2}}{%
dt_{f}},\text{ }  \label{M3.6}
\end{equation}%
where we put, by definition, that
\begin{equation}
E_{0}^{2}:=\bar{W}^{2}-c(p^{^{\prime }})^{2},\text{ \ \ \ }E_{f}:=\bar{W}%
|_{t=t_{f}}  \label{M3.7}
\end{equation}%
where $p^{\prime }:=p-\frac{q}{c}A$ is the \textit{"shifted" \ }momentum of
the charged point particle $q$ \ in the external magnetic field $B=\nabla
\times A,$ and \ took into account that the parameter $\ R_{f}:=\
R_{f}(t)|_{t=t_{f}}\in \mathbb{R}^{3}$ and the evolution parameter $%
t_{f}=t\in \mathbb{R}$ \ corresponds to  the change of the only magnetic
field potential energy $E_{f}.$ The obtained expressions (\ref{M3.6}) and (%
\ref{M3.7}) take, evidently, into account the natural balance of energies
belonging \ to both the moving charged point particle $q$ and the ambient
magnetic field.

\section{Quantum mechanics backgrounds revisited}

We will start from relation (\ref{M2.17}) rewritten in the following form:%
\begin{equation}
E_{0}^{2}=\bar{W}-p^{2}c^{2},\text{ \ }dE_{0}/dt=0,  \label{M4.1}
\end{equation}%
and make the following canonical quantization replacements:
\begin{equation}
p\rightarrow \hat{p}:=\frac{\hbar }{i}\nabla ,\text{ \ \ }E_{0}\rightarrow
\hat{E}_{0}:=-\frac{\hbar }{i}\frac{\partial }{\partial t},\text{ \ \ }\bar{W%
}\rightarrow \bar{W}\circ ,  \label{M4.2}
\end{equation}%
where $\hat{p}=\frac{\hbar }{i}\nabla $ is the standard spatial translation
generator in $\mathbb{R}^{3},$ $\hat{E}_{0}=-\frac{\hbar }{i}\frac{\partial
}{\partial t}$ is the standard time translation operator along the real time
axis $\mathbb{R}$ and $\bar{W}\circ $ is a usual scalar multiplication
operator on the function $\bar{W}:\mathbb{R}^{3}\rightarrow \mathbb{R},$ \
all acting in the Hilbert space $\mathcal{H}:=$ $L_{2}(\mathbb{R}^{3};%
\mathbb{C})$ with the standard scalar product $(\cdot ,\cdot ).$ \ As an
elementary result of this replacement we can easily write, following the
quantization recipes similar to those from \cite{Di,AB,BS,BjD}, that the
observable squared energy $E_{0}^{2}$ is the average value
\begin{equation}
E_{0}^{2}:=(\hat{E}_{0}\psi ,\hat{E}_{0}\psi )=(\psi ,(\bar{W}^{2}+\hbar
^{2}c^{2}\nabla ^{2})\psi ),  \label{M4.3}
\end{equation}%
being realized on the one-particle quantum mechanical state vector $\psi \in
\mathcal{H}.$ Taking into account representation (\ref{M4.3}) and assuming
that the Planck constant $\hbar \rightarrow 0,$ we\ now will try to
factorize the linear differential operator $\bar{W}^{2}+\hbar
^{2}c^{2}\nabla ^{2},$ well defined on a suitable dense linear subset $D(%
\hat{E}_{0})\subset \mathcal{H},$ in the following \textit{a priori }%
nonnegative canonical form:%
\begin{equation}
\bar{W}^{2}+\hbar ^{2}c^{2}\nabla ^{2}=\hat{P}^{+}\hat{P},  \label{M4.4}
\end{equation}%
where the sign\ $"+"$ \ means the standard conjugation \ operation in the
Hilbert space $\mathcal{H}.$\ It is easy to find from (\ref{M4.4}) that
\begin{equation}
\hat{P}:=\hat{S}(1+\hbar ^{2}c^{2}\bar{W}^{-1}\circ \nabla ^{2}\circ \bar{W}%
^{-1})^{1/2}\bar{W}\circ   \label{M4.5}
\end{equation}%
with an arbitrary unitary operator $\hat{S}:\mathcal{H\rightarrow H},$
satisfying the usual condition $\hat{S}^{+}\hat{S}=1.$ As a result, we
obtain from (\ref{M4.3}) and (\ref{M4.4}) that
\begin{equation}
(\hat{E}_{0}\psi ,\hat{E}_{0}\psi )=(\psi ,\hat{P}^{+}\hat{P}\psi )=(\hat{P}%
\psi ,\hat{P}\psi ).  \label{M4.6}
\end{equation}%
Thereby, based on expression (\ref{M4.6}), we can derive the following Schr%
\"{o}dinger type linear evolution equation:%
\begin{equation}
\hat{E}_{0}\psi :=i\hbar \frac{\partial \psi }{\partial t}=\hat{U}\hat{P}%
\psi   \label{M4.7}
\end{equation}%
for all $t\in \mathbb{R},$ where $\hat{U}:\mathcal{H\rightarrow H}$ is some
unitary operator to be determined later.

Let us now symbolically calculate the  operator expression (\ref{M4.5}) up
to the symbolic operator accuracy $O(\hbar ^{4}):$%
\begin{equation}
\hat{P}=\hat{S}(1+\hbar ^{2}c^{2}\bar{W}^{-1}\circ \nabla ^{2}\circ \bar{W}%
^{-1})^{1/2}\bar{W}=\hat{S}(1+\frac{\hbar ^{2}c^{2}}{2\bar{W}}\nabla
^{2}\circ \bar{W}^{-1})\bar{W}+O(\hbar ^{4}).  \label{M4.8}
\end{equation}%
Having substituted result (\ref{M4.8}) into (\ref{M4.7}) one obtains (up to
the operator accuracy $O(\hbar ^{4})$)  the next evolution equation:%
\begin{equation}
i\hbar \frac{\partial \psi }{\partial t}=\hat{U}\hat{S}(-\frac{\hbar ^{2}}{%
2(-\bar{W}/c^{2})}\nabla ^{2}+\bar{W})\psi .  \label{M4.9}
\end{equation}%
Taking into account  that the energy operator $\hat{E}_{0}=-\frac{\hbar }{i}%
\frac{\partial }{\partial t}$ is formally self-adjoint, from a physical
point of view we need to choose  in (\ref{M4.9}) that $\hat{U}:=\hat{S}^{-1},
$ as the operator $\ \hat{H}:=-\frac{\hbar ^{2}}{2(-\bar{W}/c^{2})}\nabla
^{2}+\bar{W}$ is formally self-adjoint.

Recalling now that owing to (\ref{M2.9}) and (\ref{M2.14}) the expression $-%
\bar{W}/c^{2}:=m(u)=m_{0}/\sqrt[2]{1-\frac{u^{2}}{c^{2}}}$ is the dynamical
relativistic mass of a particle, whose quantum motion is under study,
equation (\ref{M4.9}) can be finally rewritten  in the following classical
self-adjoint Schr\"{o}dinger type form:%
\begin{equation}
i\hbar \frac{\partial \psi }{\partial t}=\hat{H}\text{ }\psi :=-\frac{\hbar
^{2}}{2m(u)}\nabla ^{2}\psi +\bar{W}\psi ,  \label{M4.11}
\end{equation}%
where $\hat{H}:\mathcal{H\rightarrow H}$ is the corresponding self-adjoint
Hamiltonian operator for a relativistic point particle under an external
potential field. It is easy to observe that equation (\ref{M4.11}) in the
classical non-relativistic case when the velocity $c\rightarrow \infty $
reduces to the well-known classical Schr\"{o}dinger equation
\begin{equation}
i\hbar \frac{\partial \psi }{\partial t}=\hat{H}_{0}\text{ }\psi :=-\frac{%
\hbar ^{2}}{2m_{0}}\nabla ^{2}\psi +\bar{W}\psi ,  \label{M4.12}
\end{equation}%
derived at the beginning of the past century from completely different
points of view by the great physicists Schr\"{o}dinger, Heisenberg and Dirac.

Now we proceed to \ the similar quantization procedure of expression (\ref%
{M3.7}) for the case when the magnetic potential is not equal to zero.
Making use of the standard quantization scheme (\ref{M4.2}) we obtain, by
definition, that the average squared energy value $E_{0}^{2}$ on a suitable
quantum state vector $\psi \in \mathcal{H}$ equals
\begin{equation}
E_{0}^{2}:=(\hat{E}_{0}\psi ,\hat{E}_{0}\psi )=(\psi ,[\hat{W}^{2}-c^{2}(%
\frac{\hbar }{i}\nabla +\frac{q}{c}\hat{A})^{2}]\psi ),  \label{M4.13}
\end{equation}%
where, as before, $\hat{W}:=\bar{W}\circ $ and $\hat{A}:=\bar{A}\circ $ are
the scalar multiplication operators. The operator expression in the squared
brackets on the right-hand side of (\ref{M4.13}) can  easily be represented
in the following strongly nonnegative form:%
\begin{equation}
\hat{W}^{2}-c^{2}(\frac{\hbar }{i}\nabla +\frac{q}{c}\hat{A})^{2}:=\hat{P}%
^{+}\hat{P},  \label{M4.14}
\end{equation}%
where, as above, one can take%
\begin{equation}
\hat{P}:=[1-\hat{W}^{-1}(\frac{\hbar }{i}\nabla +\frac{q}{c}\hat{A})^{2}\hat{%
W}^{-1}]^{1/2}\hat{W}.  \label{M4.15}
\end{equation}%
Then from equation (\ref{M4.13}) and expression (\ref{M4.14}) \ we can write
down the following "magnetic" Schr\"{o}dinger type evolution equation:%
\begin{equation}
i\hbar \frac{\partial \psi }{\partial t}:=\hat{P}\psi ,  \label{M4.16}
\end{equation}%
which gives rise,  under simultaneous conditions $\hbar \rightarrow 0$ and $%
c\rightarrow \infty ,$ if the product $c\hbar =const,$ to the following
result:%
\begin{equation}
i\hbar \frac{\partial \psi }{\partial t}:=\hat{H}\psi =[\frac{1}{2m(u)}(%
\frac{\hbar }{i}\nabla +\frac{q}{c}\hat{A})^{2}+W_{0}+q\varphi ]\psi
\label{M4.17}
\end{equation}%
with accuracy $O(\hbar ^{4}).$ Equation (\ref{M4.17}) reduces to the
well-known \cite{Di} classical "magnetic" Schr\"{o}dinger type evolution
equation with accuracy $O(\hbar ^{4})$:
\begin{equation}
i\hbar \frac{\partial \psi }{\partial t}:=\hat{H}_{0}\psi =[\frac{1}{2m_{0}}(%
\frac{\hbar }{i}\nabla +\frac{q}{c}\hat{A})^{2}+W_{0}+q\varphi ]\psi ,
\label{M4.18}
\end{equation}%
where, as before, we put $m_{0}=-\bar{W}_{0}/c^{2},$ since the potential
energy relationship $|q\varphi /\bar{W}_{0}|<<1.$

The results obtained above convey very eloquently that our quasiclassical
approach to the description of the vacuum field structure, as the Planck
constant $\hbar \rightarrow 0$ and the light velocity $c\rightarrow \infty ,$
give rise to the related classical quantum Schr\"{o}dinger dynamics very
naturally, being simultaneously deeply physically motivated. It is also
worthy to mention that in all of our derivations we have used no Lorentz
invariance \ in spite of the fact that the dynamical mass \ $m=m_{0}/\sqrt[2]%
{1-\frac{u^{2}}{c^{2}}}$ \ is expressed by means of a suitable Lorentz
factor. It is, evidently, a by-product \ result of the \textit{second} field
theory principle equation (\ref{M1.2}), which is \textit{a priori }Lorentz
invariant.

To finish the discussion we make here a plausible claim that the suitable
completely relativistic quantum field theoretic analysis of our dynamical
vacuum field model via the well-known Dirac type factorization approach \cite%
{Di} will necessary shed new light on the so complicated and sophisticated
nature of the quantum world of elementary particles.

Concerning the results described above we could state that the vacuum field
theory approach of \cite{Re,PB,Br,Ca,BD,Me1,B-B} to fundamental physical
phenomena appears to be really a powerful tool in the hands of researchers,
who wish to penetrate into the hidden properties of the surrounding
Universe. As the microscopical quantum level of describing the vacuum field
matter structure is, with no doubt, very important, we see the next
challenging steps, both in understanding the backgrounds of quantum
processes within the approaches devised in \cite{Br,Re,PB} and in this work,
and in deriving new physical relationships, which will help us to explain
the Nature more deeply and adequately.

\bigskip

\bigskip

\section{Acknowledgments}

The Authors are cordially thankful to the Abdus Salam International Centre
for Theoretical Physics in Trieste, Italy, for the hospitality during their
research 2007-2008 scholarships. A.P. is, especially, grateful  to Profs.
P.I. Holod (Kyiv, UKMA), J.M. Stakhira (Lviv, NUL), B.M. Barbashov (Dubna,
JINR), Z. K{\'{a}}kol (Krak{\'{o}}w, AGH), J. S{\l }awianowski (Warsaw,
IPPT), Z. Peradzy{\'{n}}ski (Warsaw, UW) and M. B{\l }aszak (Pozna\'{n}, UP)
for fruitful discussions, useful comments and remarks. Last but not least
thanks go to Prof. O. N. Repchenko for the discussion of some controversial
vacuum field theory aspects and to Mrs. Dilys Grilli (Trieste, Publications
office, ICTP) for professional help in preparing the  manuscript for
publication.

\end{document}